# Hiding Information in Retransmissions

Wojciech Mazurczyk, Miłosz Smolarczyk, and Krzysztof Szczypiorski

Warsaw University of Technology, Institute of Telecommunications

Warsaw, Poland, 00-665, ul. Nowowiejska 15/19

**Abstract.** The paper presents a new steganographic method called RSTEG (Retransmission Steganography), which is intended for a broad class of protocols that utilises retransmission mechanisms. The main innovation of RSTEG is to not acknowledge a successfully received packet in order to intentionally invoke retransmission. The retransmitted packet carries a steganogram instead of user data in the payload field. RSTEG is presented in the broad context of network steganography, and the utilisation of RSTEG for TCP (Transport Control Protocol) retransmission mechanisms is described in detail. Simulation results are also presented with the main aim to measure and compare the steganographic bandwidth of the proposed method for different TCP retransmission mechanisms as well as to determine the influence of RSTEG on the network retransmissions level.

## 1. Classification of Network Steganography and Related Work

Communication network steganography is a method of hiding secret data in the normal data transmissions of users so that it ideally cannot be detected by third parties. Many new methods have been proposed and analysed, including those in [22], [14] and [13]. Network steganography methods may be viewed as a threat to network security, as they may be used as a tool for confidential information leakage, for example. For this reason, it is important to identify possibilities for covert communication, as knowledge of information hiding procedures may be used to develop countermeasures.

Network steganography may be classified [11] into three broad groups (Fig. 1):

- **Steganographic methods that modify packets (MP)**, including network protocol headers or payload fields.
- **Steganographic methods that modify the structure of packet streams (MS)**, for example, by affecting the order of packets, modifying inter-packet delay or introducing intentional losses.
- **Hybrid steganographic methods (HB)** that modify both the content of packets and their timing and ordering.

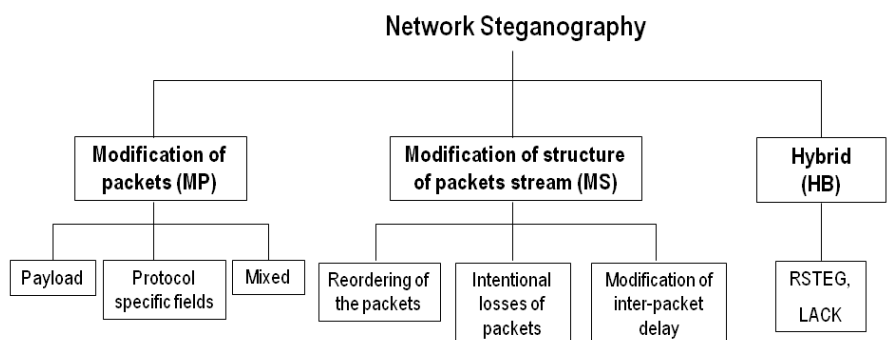

**Fig. 1** A network steganography classification

Examples of methods for each group and their characteristic features are described in Tables 1-3.

**Table 1. Examples and characteristic features of steganographic MP methods**

| MP Methods | Examples of steganographic methods | Features |
|---|---|---|
| Methods that modify protocol-specific fields | Methods based on the modification of IP, TCP, and UDP headers fields [13]. | Yield relatively high steganographic capacity. Implementation and detection is relatively straightforward. Drawbacks include potential loss of protocol functionality. |
| Methods that modify packet payload | Watermarking algorithms ([4], [2]), speech codec steganographic techniques. | Generally yield lower steganographic capacity and are harder to implement and detect. Drawbacks include potential deterioration of transmission quality, e.g., if applied to VoIP (Voice over IP). |
| Mixed techniques | HICCUPS (Hidden Communication System for Corrupted Networks, [20]). | Offer high steganographic capacity, but the implementation is more difficult than other methods due to the required low-level hardware access. For the same reason, steganalysis is harder to perform. Drawbacks include increased frame error rate. |

**Table 2. Examples and characteristic features of steganographic MS methods**

| Examples of MS methods | Features |
|---|---|
| Methods that affect the sequence order of packets [9]. | • Sender-receiver synchronisation required. |
| Methods that modify inter-packet delay [1]. | • Lower steganographic capacity and harder to detect than methods that utilise protocol-specific fields. |
| Methods that introduce intentional losses by skipping sequence numbers at the sender [17]. | • Straightforward implementation.<br>• Drawbacks include delays that may affect transmission quality. |

**Table 3. Examples and characteristic features of steganographic HB methods**

| Examples of HB methods | Features |
|---|---|
| LACK (Lost Audio PaCKets Steganography) [12].<br>RSTEG (which is presented in details in this paper). | • Modify both packets and their time dependencies.<br>• High steganographic capacity.<br>• Hard to detect.<br>• Sender-receiver synchronisation not required.<br>• Straightforward implementation.<br>• Drawbacks include a loss in connection quality. |

In the context of the above classification of network steganography methods, we propose a new hybrid method called RSTEG (Retransmission Steganography), which is intended for a broad class of protocols that utilise retransmission mechanisms. The main innovation of RSTEG is to not acknowledge a successfully received packet in order to intentionally invoke retransmission. The retransmitted packet of user data then carries a steganogram in the payload field.

Currently, there are few proposed steganographic methods that could incorporate retransmission mechanisms. Handel et al. [6] proposed a steganographic method for Ethernet CSMA/CD (Carrier Sense Multiple Access/Collision Detection), which uses a retransmission mechanism after collisions. If frame collisions occur, then a jam signal is issued, and the senders back off for a random amount of time. In order to send a single hidden bit, a back-off delay of either zero or a maximum value is used so that the hidden data rate is one bit per frame. The receiver extracts a steganogram by analysing the order of the frame arrivals after collisions.

Krätzer et al. [8] proposed a steganographic method for the 802.11 protocol, which transmits hidden information through the retransmission of frames. The sender encodes hidden data by duplicating frames transmitted to a receiver. The receiver decodes the hidden data by detecting the duplications.

The rest of the paper is dedicated to presenting the RSTEG steganographic method. Section 2 describes RSTEG in detail as well as communication scenarios in which it may be used. Performance issues involved in using the method are also discussed. In Section 3, results from an application of RSTEG to a TCP protocol simulation are presented. Section 4 concludes our work and indicates possible future research.

## 2. General Idea of RSTEG and Communication Scenarios

RSTEG can be used for all protocols that utilise retransmissions at different layers of OSI RM. A generic retransmission mechanism based on timeouts is presented in Fig. 2. RSTEG may be applied also to other retransmission mechanisms in TCP, such as FR/R (Fast Retransmit and Recovery) [18] or SACK (Selective Acknowledgement) [10].

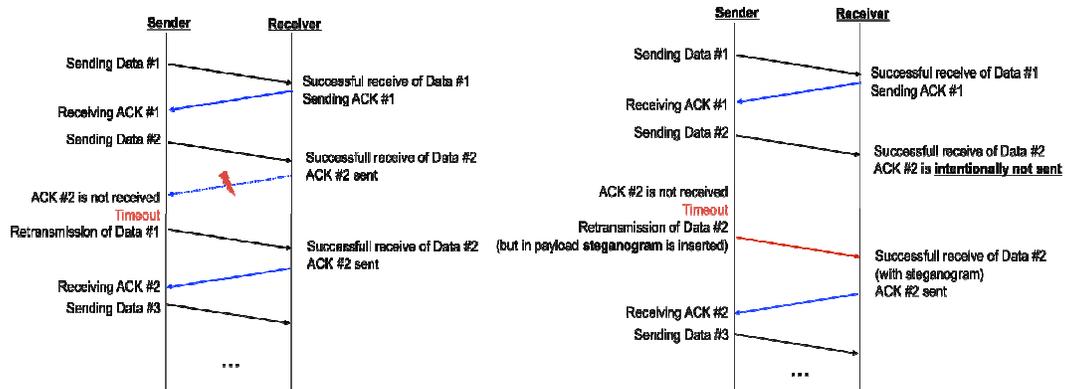

**Fig. 2.** Generic retransmission mechanism based on timeouts (left); RSTEG (right)

In a simplified situation, a typical protocol that uses a retransmission mechanism based on timeouts obligates a receiver to acknowledge each received packet. When the packet is not successfully received, no acknowledgment is sent after the timeout expires, and so the packet is retransmitted (Fig. 2).

As mentioned in Section 1, RSTEG uses a retransmission mechanism to exchange steganograms. Both a sender and a receiver are aware of the steganographic procedure. They reliably exchange packets during their connection; that is, they transfer a file. At some point during the connection after successfully receiving a packet, the receiver intentionally does not issue an acknowledgment message. In a normal situation, a sender is obligated to retransmit the lost packet when the timeframe within which packet acknowledgement should have been received expires. In the context of RSTEG, a sender replaces original payload with a steganogram instead of sending the same packet again. When the retransmitted packet reaches the receiver, he/she can then extract hidden information (Fig. 2).

Four possible hidden communication scenarios may be considered in the context of RSTEG (Fig. 3). Note that for few scenarios presented in Fig.3 packet sender and packet receiver are not taking part in hidden communication. Only a part of their communication path is utilised by intermediate nodes, which are SS (Steganogram Sender) and SR (Steganogram Receiver).

Scenario (1) is most common: the sender, who is *Steganogram Sender* (SS), and the receiver, who is the *Steganogram Receiver* (SR), engage in a connection and simultaneously exchange steganograms. The conversation path is the same as the hidden data path. RSTEG for this scenario works as follows:
(1-1)   End-to-end connection is established between sender and receiver, and the packets are exchanged.
(1-2)   At some point, the receiver does not acknowledge a successfully acquired packet.
(1-3)   After the retransmission timer expires, the packet is retransmitted, and in its payload, a steganogram is inserted.
(1-4)   The receiver is able to distinguish a retransmitted packet, so when it reaches the receiver, he/she extracts a steganogram.

In the next three scenarios (2-4 in Fig. 3), only part of the connected end-to-end path is used for hidden communication as a result of actions undertaken by intermediate nodes; the sender and/or receiver are, in principle, unaware of the steganographic data exchange.

In scenario (2), one intermediate node is involved in hidden communication with the original packet sender (SS). The steganographic procedure for this scenario works as follows:

(2-1)    While the connection lasts, one packet is selected by the sender and is marked for hidden communication.
(2-2)    When modified packet reaches the SR, the SR copies a payload and drops the packet. Now both the SS and SR know that the retransmission of this packet will be used for covert communication.
(2-3)    When the retransmission timeout expires, the packet is retransmitted by the sender, and its original payload is replaced with a steganogram.
(2-4)    When the modified retransmitted packet reaches the SR, the SR extracts a steganogram and inserts the original payload that was copied earlier and then sends it to the receiver.

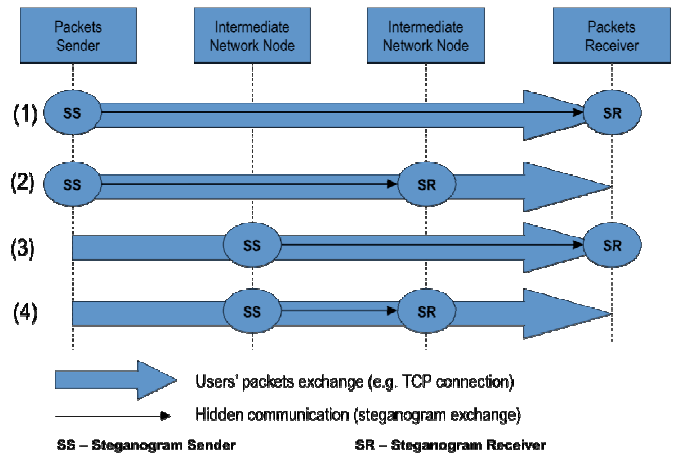

**Fig. 3** Hidden communication scenarios for RSTEG

In scenario (3), there is also one intermediate node involved in hidden communication (the SS), and the SR is located in the receiver. The steganographic procedure for this scenario works as follows:

(3-1)    While the connection lasts, one packet is selected by the intermediate node (SR) and is marked for hidden communication.
(3-2)    When the packet successfully reaches the receiver (SR), the SR intentionally does not issue an acknowledgement.
(3-3)    When the retransmission timeout expires, the packet is retransmitted by the sender.
(3-4)    When the retransmitted packet reaches SS, its payload is replaced with a steganogram.
(3-5)    When the modified, retransmitted packet reaches SR, the SR extracts a steganogram.

In scenario (4), two intermediate nodes are involved in hidden communication and are utilising existing end-to-end connection between sender and receiver. RSTEG for this scenario works as follows:

(4-1)    While the connection lasts, one packet is selected by the SS and is marked for hidden communication.
(4-2)    When the modified packet reaches the SR, the SR copies the payload and drops the packet. Now both the SS and SR know that retransmission of this packet will be used for covert communication.
(4-3)    When the retransmission timeout expires, the packet is retransmitted by the sender.
(4-4)    When retransmitted packet reaches the SS, its payload is replaced with steganogram.
(4-5)    When the modified retransmitted packet reaches the SR, the SR extracts the steganogram and inserts the original payload that was copied earlier and sends it to the receiver.

Of the above scenarios, scenario (1) is easiest to implement; scenarios (2)-(4) require control over the intermediate node used for hidden communication and that all packets traverse through it during connection. On the other hand scenarios (2), (3) and, in particular, (4) are harder to detect than (1). The typical location of the node used for steganalysis is near the sender or receiver of the packets. Thus, in scenarios in which only part of the communication path is used, it may be harder to uncover.

The performance of RSTEG depends on many factors, such as the details of the communication procedure (in particular the size of the packet payload, the rate at which segments are generated, and so on). No real-world steganographic method is perfect; whatever the method, the hidden information can be potentially discovered. In general, the more hidden information is inserted into the data stream, the greater the chance that it will be detected, for example, by scanning the data flow or by some other steganalysis methods.

Moreover, the more packets that are used to send covert data, the higher the retransmission rate will be, which allows easier detection. That is why the procedure of inserting hidden data has to be carefully chosen and controlled in order to minimise the chance of detecting inserted data.

Additionally, packet losses introduced by the network must be carefully monitored. Because RSTEG uses legitimate traffic, it thus increases overall packet losses. To ensure that the total packet loss introduced by the network and by RSTEG is not too high if compared with other connections in the same network, the level of the retransmissions used for steganographic purposes must be controlled and dynamically adapted.

## 3. RSTEG in TCP Protocol: Functioning, Detection and Experimental Results

Applying RSTEG to TCP is the natural choice for IP networks, as a vast amount of Internet traffic (about 80-90%) is based on this protocol. For TCP, the following retransmission mechanisms are defined:

- **RTO** (Retransmission Timeouts) [15] in which segment loss detection is based on RTO timer expiration. Results from [16] show that 60-88% of all retransmissions on the Internet were caused by RTO mechanism. In RTO, a segment is considered lost if the receiver does not receive an acknowledgement segment (ACK) after the specified period of time, after which it is retransmitted. The RTO timer value varies in TCP implementation across different operating systems, and it depends mainly on RTT (Round Trip Time) and its variation. If the RTO timer is set to too low of a value, it may cause too many spurious retransmissions; otherwise, the sender will be waiting too long to retransmit a lost segment, which may cause throughput decrease.
- **FR/R** (Fast Retransmit/Recovery) is based on detecting duplicate ACKs (that is, ACKs with the same acknowledgement number). A receiver acknowledges all segments delivered in order. When segments arrive out of order, the receiver must not increase the acknowledgement number so as to avoid data gaps but instead sends ACKs with unchanged acknowledgement number values, which are called duplicate ACKs (dupACKs). Usually, a segment is considered lost after the receipt of three duplicate ACKs. Issuing duplicate ACKs by the receiver is often a result of out-of-order segment delivery. If the number of duplicate ACKs that triggers retransmission is too small, it can cause too many retransmissions and can degrade network performance.
- **SACK** (Selective Acknowledgement) is based on Fast Retransmit/Recovery. It uses an extended ACK option that contains blocks edges to deduce which received blocks of data are non-contiguous. When retransmission is triggered, only missing segments are retransmitted. This feature of SACK decreases network load.

### 3.1 RSTEG Insertion and Extracting Procedures for TCP

The intentional retransmissions due to RSTEG should be kept at a reasonable level to avoid detection. To achieve this goal, it is necessary to determine the average number of natural retransmissions in TCP-based Internet traffic as well as to know how intentional retransmissions affect the network retransmission rate. Usually network retransmissions are caused by network overload, excessive delays or reordering of packets [16], and their number is estimated to account for up to 7% of all Internet traffic [16, 7, 3].

RSTEG can be applied to all retransmission mechanisms presented above. It requires modification to both a sender and a receiver. A sender should control the insertion procedure and decide when a receiver should invoke a retransmission. The sender is also responsible to keep the number of retransmissions at a non-suspicious level. The receiver's role is to detect when the sender indicates that intentional retransmission should be triggered. Then, when the retransmitted segment arrives, the receiver should be able to extract the steganogram.

The sender must be able to mark segments selected for hidden communication (that is, retransmission request segments) so the receiver would know which segments retransmissions should be invoked and which segments will contain steganograms. However, marked TCP segment should not differ from those sent during a connection. The following procedure for marking sender segments is proposed. Let us assume that the sender and receiver share a secret Steg-Key (*SK*). For each fragment chosen for steganographic communication, the following hash function (*H*) is used to calculate the Identifying Sequence (*IS*):

$$IS = H(SK \parallel Sequence\ Number \parallel TCP\ Checksum \parallel CB) \qquad (3\text{-}1)$$

Note that *Sequence Number* and *TCP Checksum* denote values from the chosen TCP header fields in segments, $\parallel$ is the bits concatenation function, and *CB* is a control bit that allows the receiver to distinguish a retransmission request segment from a segment with a steganogram. For every TCP segment used for hidden communications, the resulting *IS* will have different value due to the variety of values in the *Sequence Number* and *TCP Checksum* header fields. All *IS* bits (or only selected ones) are distributed by the sender across a segment's payload field in a predefined manner. The receiver must analyse each incoming segment; based on *SK* and values from the TCP header, the receiver calculates two values of *IS*, namely, one with *CB* = 1 and one with *CB* = 0. Then the receiver checks if and which *IS* is present inside the received segment.

Problems may arise when the segment that informs the receiver of a necessity to invoke an intentional retransmission (which contains user data together with the *IS*) is lost due to network conditions. In that case, a normal retransmission is triggered, and the receiver is not aware that the segment with hidden data will be sent. However, in this case, the sender believes that retransmission was invoked intentionally by the receiver, and so he/she issues the segment with steganogram and the *IS*. In this scenario, user data will be lost, and the cover connection may be disturbed.

In order to address the situation in which the receiver reads a segment with an unexpected steganogram, the receiver should not acknowledge reception of this segment until he/she receives the segment with user data. When the ACK is not sent to the sender, another retransmission is invoked. The sender is aware of the data delivery failure, but he/she does not know which segment to retransmit, so he/she first issues a segment with user data. If delivery confirmation is still missing, then the segment with steganogram is sent. The situation continues until the sender receives the correct ACK. This mechanism of correcting steganogram network losses is illustrated in Fig. 4.

For example, consider the scenario in which we invoke 0.5% intentional retransmissions. If 5% is lost, it means that the above-described mechanism will take place only for 0.025% of steganogram segments, and thus it will be used rarely.

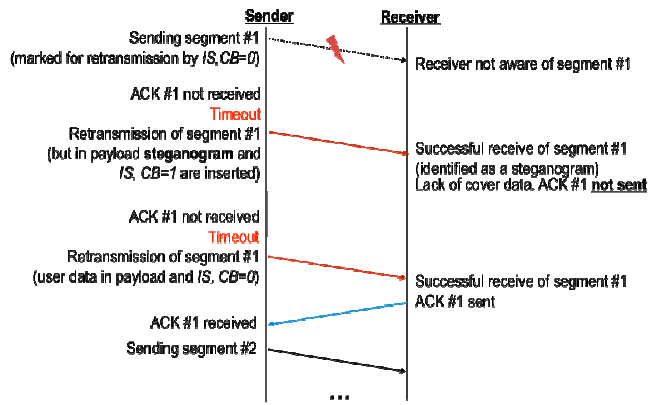

**Fig. 4** RTO-based RSTEG segment recovery example

The above RSTEG may be applied to the retransmission mechanisms presented above as follows:

- **RTO-based RSTEG:** The sender marks a segment selected for hidden communication by distributing the *IS* across its payload. After successful segment delivery, the receiver does not issue an ACK message. When the RTO timer expires, the sender sends a steganogram inside the retransmitted segment's payload (see Fig. 2). The receiver extracts the steganogram and sends the appropriate acknowledgement.
- **FR/R-based RSTEG:** The sender marks the segment selected for hidden communication by distributing the *IS* across its payload. After successful segment delivery, the receiver starts to issue duplicate ACKs to trigger retransmission. When the ACK counter at the sender side exceeds specified value, the segment is retransmitted (see Fig. 5). Payload of the retransmitted segment contains a steganogram. The receiver extracts the steganogram and sends an appropriate acknowledgement.

- **SACK based RSTEG:** The scenario is exactly the same as FR/R, but in the case of SACK, it is possible that many segments are retransmitted because of potential non-contiguous data delivery.

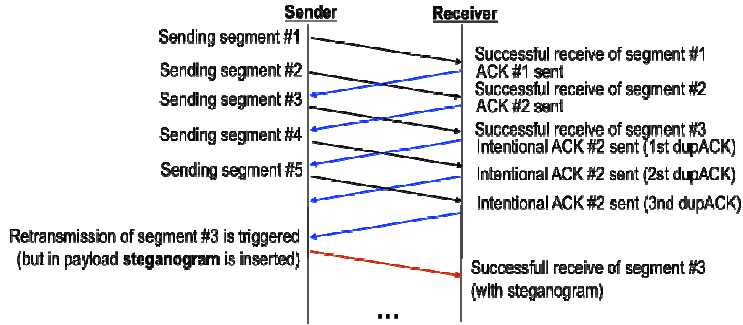

**Fig. 5** FR/R-based RSTEG

### 3.2 An Experimental Evaluation of the Influence of RSTEG on TCP Connections

Simulations were generated using ns-2 Simulator ver. 2.33 [21] with the following modifications. The adaptation of ns-2 Simulator to RSTEG required only modifications of the receiver. The receiving functionality of the segments was modified to intentionally not issue ACKs (in the case of RTO) or to not increase the acknowledgement number (in the cases of FR/R and SACK). The decision regarding which segment would be treated as lost is made randomly according to a parameter that specifies the intentional retransmissions frequency.

The network topology was matched to fit Internet traffic retransmission statistics. Simulation scenario consists of two traffic sources (TCP and UDP) and the bottleneck link between intermediate devices such as routers (see Fig. 6). Each traffic source is connected with a 10 Mbps link to the intermediate device. The receiver is also connected to its router with 10 Mbps link. The UDP traffic source and the bandwidth of the link between intermediate devices ($X$) are chosen to introduce certain network retransmission probabilities ($NR_P$); due to network overload, $NR_P$ is about 3 or 5%. Table 4 summarises the bandwidths of the bottleneck links that are used for simulation purposes.

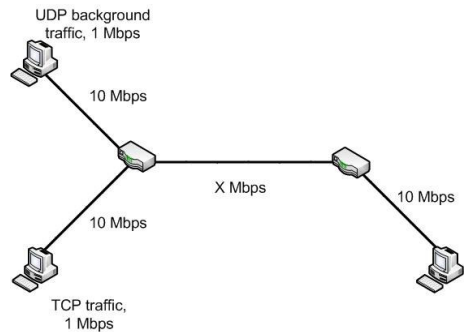

**Fig. 6** RSTEG simulation scenario

**Table 4.** The chosen bandwidth for bottleneck link ($X$) for different TCP retransmission mechanisms to achieve 3% and 5% $NR_P$

| $NR_P$ / TCP retrans. | RTO | FR/R | SACK |
|---|---|---|---|
| 3% | 1.985 Mbps | 1.985 Mbps | 1.985 Mbps |
| 5% | 1.8 Mbps | 1.8 Mbps | 1.9 Mbps |

The simulation results are based on comparing retransmissions for a network with RSTEG applied to TCP traffic as well as for a network without RSTEG retransmissions. Network traffic was measured for 9 minutes, starting after 1 minute from the beginning of the simulation. The RSTEG intentional retransmission probability ($IR_P$) was changed from 0 to 5% with intermediary steps at 0%, 0.5%, 1%, 2%, 3%, 4% and 5%.

In the above simulation scenario, two parameters were measured for RSTEG:

- **Steganographic Bandwidth ($S_B$)** is defined as the amount of the steganogram transmitted using RSTEG during one second [Bps]. For different retransmission mechanisms in the TCP protocol, this parameter can be used to estimate which mechanism yields the highest $S_B$ and is most suitable from a RSTEG utilisation point of view. $S_B$ depends mainly on the size of the segment and the number of intentional retransmissions invoked, and so it may be expressed as

$$S_B = \frac{N_S \cdot S_S}{T} \quad [Bps] \qquad (3\text{-}2)$$

where
  $N_S$ – the number of segments used for hidden communication
  $S_S$ – the size of segment payload
  T – the duration of the connection

- **Retransmissions Difference ($R_D$)** is defined as the difference between retransmissions in a network after applying RSTEG and in a network before applying RSTEG. This parameter can be used to estimate the influence that RSTEG has on the TCP retransmissions rate. Thus, it can illustrate how to choose the correct intentional retransmission probability to limit the risk of detection. For example, if the network retransmission probability is 5%, 1% of intentional retransmissions are introduced by RSTEG, which causes the overall retransmission rate to increase to 7%, with $R_D = 2\%$.

The results for TCP retransmission mechanisms when $NR_P = 3\%$ and $NR_P = 5\%$ are presented in Fig. 7-10.

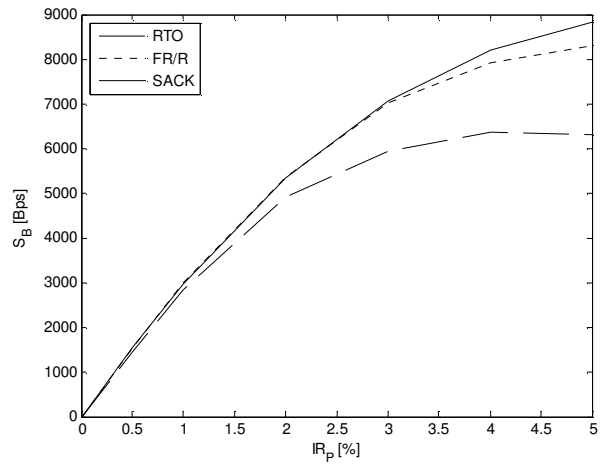

**Fig. 7** $S_B$ for TCP retransmission mechanisms when $NR_P = 3\%$ and $IR_P$ varies

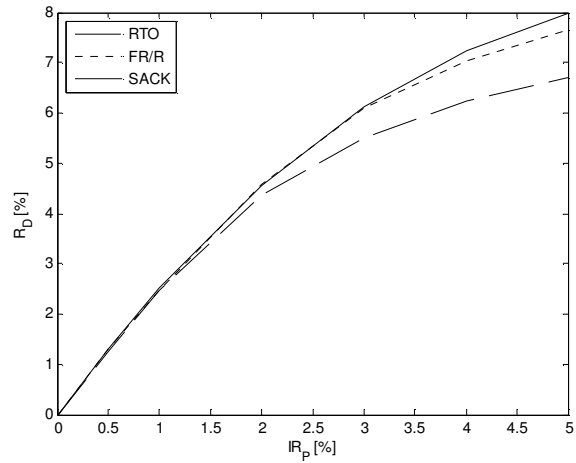

**Fig. 8** $R_D$ for TCP retransmission mechanisms when $NR_P = 3\%$ and $IR_P$ varies

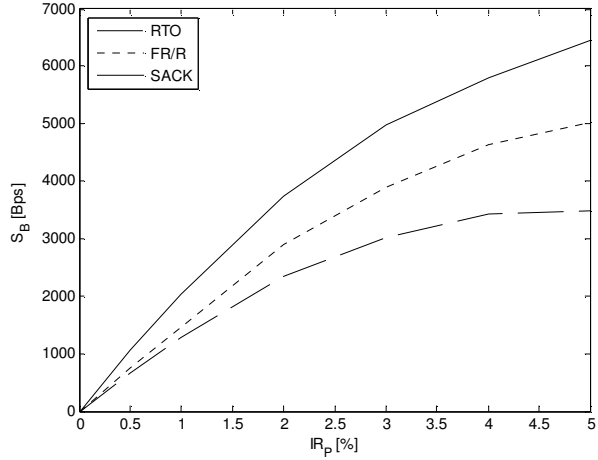

**Fig. 9** $S_B$ for TCP retransmission mechanisms when $NR_P = 5\%$ and $IR_P$ varies

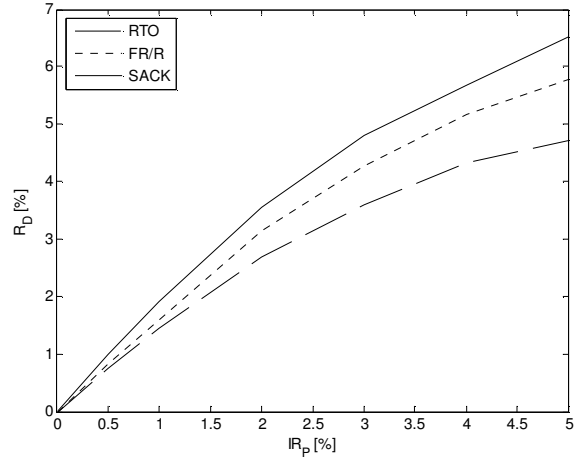

**Fig. 10** $R_D$ for TCP retransmission mechanisms when $NR_P = 5\%$ and $IR_P$ varies

Tables 5 and 6 summarise the simulation results.

**Table 5.** Simulation results when $NR_P = 3\%$

| | *RTO* | | | | *FR/R* | | | | *SACK* | | | |
|---|---|---|---|---|---|---|---|---|---|---|---|---|
| $IR_P$ [%] | $S_B$ [Bps] | $\sigma_{SB}$ | $R_D$ [%] | $\sigma_{RD}$ | $S_B$ [Bps] | $\sigma_{SB}$ | $R_D$ [%] | $\sigma_{RD}$ | $S_B$ [Bps] | $\sigma_{SB}$ | $R_D$ [%] | $\sigma_{RD}$ |
| 0.5 | 1 454 | 112.5 | 1.25 | 0.0971 | 1 530 | 92.8 | 1.25 | 0.1292 | 1 530 | 92.8 | 1.29 | 0.0778 |
| 1.0 | 2 821 | 164.3 | 2.45 | 0.1356 | 2 999 | 141.1 | 2.54 | 0.1302 | 2 999 | 141.1 | 2.54 | 0.1183 |
| 2.0 | 4 802 | 183.4 | 4.26 | 0.1503 | 5 395 | 171.3 | 4.67 | 0.1773 | 5 395 | 171.3 | 4.62 | 0.1445 |
| 3.0 | 5 982 | 96.4 | 5.54 | 0.0754 | 7 113 | 106.6 | 6.12 | 0.1384 | 7 113 | 106.6 | 6.17 | 0.0896 |
| 4.0 | 6 306 | 100.7 | 6.21 | 0.0911 | 8 128 | 157.3 | 7.03 | 0.1119 | 8 128 | 157.3 | 7.18 | 0.1355 |
| 5 | 6 320 | 81.5 | 6.72 | 0.0800 | 8 865 | 62.0 | 7.73 | 0.0830 | 8 865 | 62.0 | 8.07 | 0.0754 |

**Table 6.** Simulation results when $NR_P = 5\%$

| | *RTO* | | | | *FR/R* | | | | *SACK* | | | |
|---|---|---|---|---|---|---|---|---|---|---|---|---|
| $IR_P$ [%] | $S_B$ [Bps] | $\sigma_{SB}$ | $R_D$ [%] | $\sigma_{RD}$ | $S_B$ [Bps] | $\sigma_{SB}$ | $R_D$ [%] | $\sigma_{RD}$ | $S_B$ [Bps] | $\sigma_{SB}$ | $R_D$ [%] | $\sigma_{RD}$ |
| 0.5 | 5 457 | 677 | 0.77 | 0.0680 | 5 474 | 694 | 0.75 | 0.0666 | 6 119 | 1 020 | 0.96 | 0.0938 |
| 1.0 | 6 068 | 1 288 | 1.46 | 0.0929 | 6 277 | 1 497 | 1.61 | 0.0736 | 7 119 | 2 020 | 1.90 | 0.1019 |
| 2.0 | 7 169 | 2 389 | 2.75 | 0.1186 | 7 699 | 2 919 | 3.15 | 0.1473 | 8 726 | 3 627 | 3.44 | 0.1092 |
| 3.0 | 7 848 | 3 068 | 3.62 | 0.1014 | 8 692 | 3 912 | 4.28 | 0.0982 | 9 998 | 4 899 | 4.71 | 0.1323 |
| 4.0 | 8 173 | 3 393 | 4.24 | 0.0881 | 9 406 | 4 626 | 5.14 | 0.1390 | 10 886 | 5 787 | 5.67 | 0.1115 |
| 5 | 8 304 | 3 524 | 4.74 | 0.1216 | 9 863 | 5 083 | 5.81 | 0.1101 | 11 549 | 6 450 | 6.50 | 0.0929 |

Based on the results presented above, one can conclude that for low intentional retransmission probability values (0 – 0.5% for $NR_P$ = 5% and 0 – 1% for $NR_P$ = 5%), the resulting $S_B$ values for all retransmission mechanisms are similar, and therefore, it is not important which of the retransmission mechanisms (that is, RTO, FR/R or SACK) is used. The higher $IR_P$ is, the greater is the difference in steganographic bandwidth. It is not surprising that RSTEG based on SACK and FR/R mechanisms yield higher steganographic bandwidth than RTO-based RSTEG, as the former are more effective retransmission mechanisms. That is, under the same $IR_P$, they achieve greater $S_B$. However, higher steganographic bandwidth for RSTEG based on SACK and FR/R mechanisms increases the retransmission difference values in comparison to RTO-based RSTEG. This may increase the likelihood of detection of RSTEG. Thus, retransmission mechanisms for which $R_D$ values are lower are favourable in terms of steganalysis. RTO-based RSTEG achieved the lowest steganographic bandwidth but simultaneously introduced the lowest $R_D$. Considering this analysis and knowing that RTO is the most frequent retransmission mechanism used for TCP on the Internet (60-88%) suggests that RTO-based RSTEG is a favourable choice for TCP protocol if the risk of disclosure must be minimized. If detection issues are omitted, SACK-based RSTEG should be chosen to maximise the amount of steganogram that is sent.

Regarding RTO-based RSTEG and its appropriateness based on TCP protocol, Fig. 10 and 11 present a comparison of $S_B$ and $R_D$ when $IR_P$ = 3% and $IR_P$ = 5%.

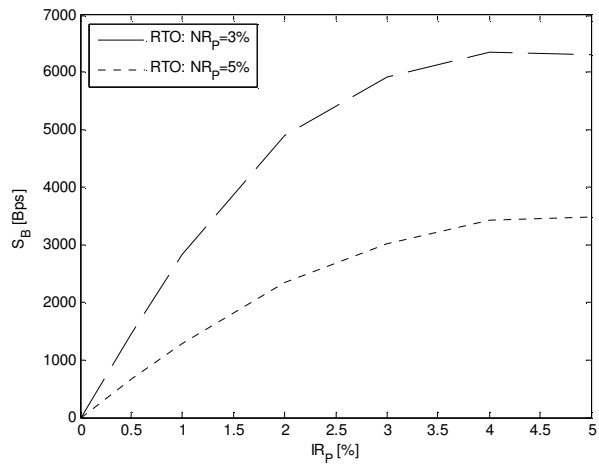

**Fig. 11** $S_B$ for RTO-based RSTEG as $IR_P$ varies

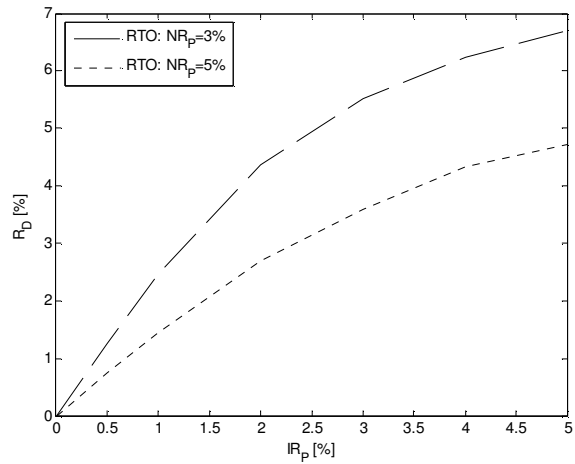

**Fig. 12** $R_D$ for RTO-based RSTEG as $IR_P$ varies

Results from Fig. 11 and 12 show that an increase in the number of retransmissions introduced in a network lowers the influence that RSTEG has on network retransmissions. That is, they are harder to detect, although the steganographic bandwidth is lower. An increase in network retransmissions means that it is easier to hide intentional retransmissions among unintentional retransmissions.

**3.3 RSTEG Detection**

Retransmissions in IP networks are a "natural phenomenon", and so intentional retransmissions introduced by RSTEG are not easy to detect if they are kept at a reasonable level. The experimental results presented here show that RTO-based RSTEG is a favourable TCP retransmission mechanism in terms of steganalysis. Moreover, if the sender can observe the average retransmission rate in a network, then he/she can also choose an $IR_P$ so as to limit the risk of detection.

One possible detection method is statistical steganalysis based on the network retransmission rate. If for certain TCP connections, the retransmission rate is significantly higher than for others, then potential usage of RSTEG may be detected. Such a steganalysis method involves the monitoring of TCP retransmission rates for all connections in a sub-network.

However, there is a solution that makes the steganalysis of RSTEG as applied to TCP protocol easier to perform. The proposed steganalysis method may be implemented with a passive warden [5] (or some other network node responsible for steganography usage detection). Passive warden must be able to monitor all the TCP traffic and for each TCP connection it must store sent segments for the given period of time, which depends on the retransmission timer i.e. passive warden must store the segment until it is acknowledged by the receiver so the retransmission is not possible any more. When there is a retransmission issued, passive warden compares originally sent segment with retransmitted one and if the payload differs RSTEG is detected and the segment is dropped. However, it should be noted that there may be serious performance issues involved if passive warden monitors all the TCP connections and must store a large number of the segments.

On the other hand, it must be noted that based on results presented in [19] up to 0.09 % (1 in 1100) of TCP segments may be corrupted due to network delivery. As a result, an imperfect copy of a segment may be sent to the receiver. After reception of the invalid segment, verification is performed based on the value in the *TCP Checksum* field, and the need to retransmit is signalled to the sender. Thus, in this scenario, the original segment and the retransmitted one will differ from each other. Occurrences of this effect in IP networks mask the use of RSTEG. Thus, the steganalysis methods described above may fail, because the warden will drop retransmitted segments when differences among segments are discovered, and as a result, user data will be lost.

It is worth noting that even for the low rates of intentional retransmission (0.09%) that are required to mask RSTEG, if we assume that the TCP segments are generated at a rate of 200 segments/s, with the connection lasting 5 minutes and the segment's payload size being 1000 bytes, then this results in $S_B$ = 180 Bps, which is a rather high bandwidth, considering the other steganographic methods presented in Section 1.

To summarise, measures to detect RSTEG have been proposed and can be utilised, but if the rate of intentional retransmissions is very low, then the detection of hidden communications may be difficult.

## 4. Conclusions and Future Work

RSTEG is a hybrid network steganographic method based on the classification presented earlier in this paper. The steganographic bandwidth it can provide may be comparable for methods that modify packets only, and its bandwidth is higher than that of methods that only modify the structure of packet streams.

In this paper, we have focused on presenting the framework guiding this steganographic method and have showed how it may be applied and detected in the context of TCP protocol, which may be useful in developing detection measures. A more detailed evaluation of RSTEG performance for other protocols with retransmissions and in other layers of the TCP/IP stack is needed.

The simulation results show that in order to minimise the risk of detection, RTO-based retransmissions should be used by RSTEG, and intentional retransmissions should be kept to a reasonable level. However, in order to maximise steganographic bandwidth, SACK-based RSTEG is more appropriate.

Application of RSTEG to TCP protocol is a logical choice for IP networks, but as shown in this paper, it can be detected, especially if intentional retransmissions are issued excessively. Nevertheless, RSTEG can be also used for other protocols that utilise retransmission mechanisms, in particular for wireless networks. We believe that RSTEG in this environment may be harder to detect; however, this claim requires a more detailed analysis. Analytical and experimental results concerning this issue will be presented by the authors in forthcoming papers.